\documentclass[twocolumn,prl,aps,showpacs]{revtex4}
\usepackage{graphicx}
\def\Ref#1{(\ref{#1})}
\newcommand{\be}{\begin{equation}}
\newcommand{\ee}{\end{equation}}
\newcommand{\bn}{\begin{eqnarray}}
\newcommand{\en}{\end{eqnarray}}
\usepackage[russian]{babel}
\begin{document}
\draft

\title{A simple measure of memory for dynamical processes\\
described by the generalized Langevin equation}

\author{\firstname{Anatolii~V.}~\surname{Mokshin}}
\email{mav@theory.kazan-spu.ru} \affiliation{Department of
Physics, Kazan State Pedagogical University, 420021 Kazan, Russia
}

\author{\firstname{Renat~M.}~\surname{Yulmetyev}}
\affiliation{Department of Physics, Kazan State Pedagogical
University, 420021 Kazan, Russia}

\author{\firstname{Peter}~\surname{H\"anggi}}
\affiliation{Department of Physics, University of Augsburg,
D-86135 Augsburg, Germany}

\date{\today}
\begin{abstract}
Memory effects are a key feature in the description of the dynamical
systems governed by the generalized Langevin equation, which
presents an exact reformulation of the equation of motion. A simple
measure for the estimation of memory effects is introduced within
the framework of this description. Numerical calculations of the
suggested measure and the analysis of memory effects are also
applied for various model physical systems as well as for the
phenomena of ``long time tails'' and anomalous diffusion.
\end{abstract}
\pacs{05.40.-a, 02.50.Ey, 05.60.-k} \maketitle


An abundant number of different problems involving molecular
motions in condensed matters can be formulated in terms of the
generalized Langevin equation (GLE). The latter can be exactly
received in the context of the Zwanzig-Mori projection operator
technique \cite{Zwanzig_book}, as well as within the framework of
the recurrent relation approach \cite{Lee1} from the equation of
motion. As known, this approach is used in the description of
dynamical phenomena, ordinary and anomalous (such as anomalous
diffusion) transport in physical, chemical and even biophysical
complex systems \cite{Zwanzig_book,Hanggi}.

One of the key features of the GLE is the fact that it contains an
aftereffect function, termed a memory function. If the memory
function is delta-function correlated (``a white noise''), the GLE
is reduced to the ordinary Langevin equation corresponding to the
system without memory, and the time correlation function (TCF)
related to the momentum degree of freedom has simple exponential
relaxation. So, a formalism based on GLE (and/or likewise, on the
corresponding generalized Fokker-Planck equation) inherently
contains  \textit{memory effects}, which characterize the system of
interest \cite{GHT}.

Memory effects can appear in the velocity auto-correlation function
(VACF) either through the presence of an oscillatory behavior or by
means of slowly decreasing correlations. Thus, the memory character
of the  dynamics of liquids related to influence of the cage formed
by nearest surroundings on the particle movement (the so-called
``cage effects'') belongs to the first case, whereas dynamical
processes related to anomalous diffusion and long time tails in the
behavior of TCF's correspond to the second situation. It is
necessary to note that in the similar treatment the notion of
``memory'' is not completely definite. In particular, it is found
that non-exponential relaxation \cite{Lee0} can be observed in
glasses, simple fluids and supercooled liquids
\cite{Balucani_Zoppi}, liquid crystals \cite{Ben}, plasmas
\cite{Fer}, frustrated lattice gases \cite{Vain}, proteins
\cite{Peyr}, disordered vortex lattice in superconductors
\cite{Bouch}. However, in concordance with GLE treatment, the
process with memory effects constitutes \textit{any} dynamical
process, whose TCF decays, at long times, according to a
non-exponential law. As a result, the following questions arise. To
what degree is the studied process not Markovian? Correspondingly,
how large is the difference between the relaxation with memory and
an exponential relaxation? How strong are the space and time
non-locality effects in the system? The situation would become more
clear via the characterization by an adequate quantity \cite{HT},
which would provide \textit{a numerical measure of manifest memory
effects}. In this Letter we present such a measure: it is obtained
by comparing the time scales of the decay of the VACF and its
corresponding memory function.

Thus, the  GLE of Mori-type can be written as \cite{Hanggi,Lee1}:
\be m\frac{d}{dt}v_{\alpha}(t)=-m\omega^{(2)}\int_{0}^{t}
M_1(t-\tau)v_{\alpha}(\tau) d\tau + F(t). \label{GLE_1} \ee where
$\omega^{(2)}= \langle |\mathcal{L} v_{\alpha}|^2 \rangle/\langle
|v_{\alpha}|^2\rangle$ is the second frequency moment of VACF with
the Liouville operator $\mathcal{L}$, and the normalized memory
function $M_1(t)$ can be related to the stochastic force $F(t)$ by
means of the fluctuation-dissipation theorem, $\langle F(t) F(0)
\rangle = m k_B T \omega^{(2)} M_1(t)$. \noindent Here $k_{B}$, $T$
and $m$ are the Boltzmann constant, the temperature and the mass of
a particle, respectively. Multiplying Eq. \Ref{GLE_1} by
$v_{\alpha}(0)/\langle |v_{\alpha}(0)|^2 \rangle$ and performing an
appropriate ensemble average $\langle \ldots\rangle$, one obtains
the memory equation for the VACF, i.e.:  \be
\frac{da(t)}{dt}=-\omega^{(2)} \int_{0}^{t}
M_{1}(\tau)a(t-\tau)d\tau,\  \langle v_{\alpha}(0)F(t)\rangle=0,
\label{GLE_2} \ee here the VACF $a(t)=\langle v_{\alpha}(0)
v_{\alpha}(t) \rangle/\langle v_{\alpha}(0)^{2} \rangle$ \noindent
appears, which is related to the diffusion constant by the integral
relation \cite{Boon} $D=(k_{B}T/m)\int_{0}^{\infty}\;a(t)dt$.

Next we introduce a dimensionless parameter which
characterizes memory effects for the relaxation process related to
the velocity degree of freedom \be \delta =
\frac{\tau_{a}^2}{\tau_{M}^2}, \label{parameter} \ee \be
\tau_{a}^2=\left |\int_0^{\infty} t a(t) dt \right |, \ \
\tau_{M}^2=\left |\int_0^{\infty} t M_1(t) dt \right |,
\label{freq} \ee where $\tau_{a}^2$ and $\tau_{M}^2$ are the
squared characteristic relaxation scales of the VACF and its
corresponding memory function. Note that $0 \leq \delta < \infty$.
Obviously, the situation $\tau_{a}^2 \gg \tau_{M}^2$ corresponds
to a memoryless  behavior with a fast-decaying memory. Then, in
accordance with Eq. \Ref{parameter} we find that $\delta \to
\infty$. For processes exhibiting  strong, pronounced memory
effects, $\tau_{a}^2 \ll \tau_{M}^2$, this parameter thus
approaches $\delta \to 0$ \cite{sm}.

Eqs. \Ref{freq} can be rewritten in the terms of Laplace transforms
of corresponding correlation functions,
$\widetilde{f}(z)=\int_0^{\infty} e^{-zt} f(t)dt$, in following
forms: \be \tau_a^2=\left |\lim_{z \to 0} \left ( -
\frac{\partial\widetilde{a}(z)}{\partial z} \right ) \right |, \ \ \
\tau_M^2=\left |\lim_{z \to 0} \left ( -
\frac{\partial\widetilde{M}_1(z)}{\partial z} \right ) \right |.
\label{proizv} \ee

After a Laplace transform, Eq. \Ref{GLE_2} reads: \be
\tilde{a}(z)=[z+\omega^{(2)}\widetilde{M}_1(z)]^{-1}, \label{GLE_3}
\ee which allows one to obtain \be
-\widetilde{M}_1'(z)=\frac{\widetilde{a}'(z)+\widetilde{a}(z)^2}{\omega^{(2)}\widetilde{a}(z)^2},
\ee where  $\widetilde{a}'(z)=\partial \widetilde{a}(z)/\partial z$
and $\widetilde{M}_1'(z)=\partial \widetilde{M}_1(z)/\partial z$.
Then, taking into account the relations \Ref{proizv} we recast Eq.
\Ref{parameter} as \bn \delta= \omega^{(2)}\left |\lim_{z \to 0}
\frac{\widetilde{a}'(z)
\widetilde{a}(z)^2}{\widetilde{a}'(z)+\widetilde{a}(z)^2} \right |.
\label{nM} \en Note that the knowledge of a(t)  and its  first
derivative at long times,
 i.e. for $ z \rightarrow 0$,  and $\omega^{(2)}$, are sufficient to evaluate
the measure $\delta$. The complete knowledge of the memory function
is thus not required. Moreover, the frequency moment of the VACF
$\omega^{(2)}$, and thus $\delta$, is directly related to physical
observable quantities such as the distribution function and the
potential of inter-particle interaction \cite{Resibua}.

Let us consider two contrasting situations: i) the system has a
short-range memory, and ii) the system is characterized by
physical (long-range) memory.

In the first case (\textit{the memory-free limit}) the memory
function can be presented as $M_{1}(t)=2\tau_{1}\delta(t)$. Then
Eq. \Ref{GLE_2} is reduced to the ordinary Langevin equation
\cite{Langevin}: \be m\frac{d}{dt}v_{\alpha}(t)+m\omega^{(2)}
\tau_1 v_{\alpha}(t)=F(t) \label{NLE} \ee with a simple
exponential solution for VACF: \be
a(t)=\textrm{e}^{-\omega^{(2)}\tau_{1}t}, \label{Brown} \ee which
is correct for the VACF of a free Brownian particle with the
relaxation time $\tau_{0}=(\omega^{(2)}\tau_{1})^{-1}=m/\gamma$,
$m$ and $\gamma$ are the mass and the friction coefficient,
respectively. Let us define the parameter $\delta$ for this case.
Applying the Laplace transform to the memory function and VACF
[Eq. \Ref{Brown}] for this case, and taking into account Eqs.
\Ref{proizv} one can find that $\tau_a^2=\tau_0^2$ and
$\tau_M^2=0$. Then, Eq. \Ref{parameter} yields exactly $\delta \to
\infty$.

The opposite situation is appropriate for  systems that exhibit
indefinitely large memory, i.e. the so-called \textit{strong memory
limit}. The  time dependence for this model can be presented
adequately as follows: \be M_{1}(t)=H(t)=\left\{
\begin{array}{rl}
1, &t \geq 0\\
0, &t < 0,\\
\end{array}
\right.  \label{hevisaid} \ee where $H(t)$ is the step Heavisaid
function. Introducing Eq. \Ref{hevisaid} into the convolution
integral of Eq. \Ref{GLE_2}, one obtains the following equation
\bn \frac{da(t)}{dt}=-\omega^{(2)}\int_{0}^{t} a(\tau)\; d\tau,
\nonumber \en which has the simple solution \be
a(t)=\cos(\sqrt{\omega^{(2)}}t). \label{cos} \ee The similar VACF
describes oscillations of the lattice particles in crystalline
solids. Using the Laplace transforms of Eqs. \Ref{hevisaid} and
\Ref{cos} we define from Eq. \Ref{proizv} the squares of
characteristic relaxation scales of VACF and memory function,
$\tau_a^2=1/\omega^{(2)}$ and $\tau_M^2 \to \infty$. Then, from
definition \Ref{parameter} we find $\delta \to 0$. Thus, we have
shown that the parameter $\delta$ possesses true meaning for both
the memory-free and the manifest, strong memory limit.

We can therefore state with confidence that the quantity $\delta$
sensitively detects and measures the presence of memory effects.
The offered measure of memory can be obtained theoretically,
experimentally and with the help of computer modeling if
$\widetilde{a}(z)$ [or $a(t)$] is known. We next evaluate $\delta$
for various model systems, thereby demonstrating its usefulness.

A. \textit{The Rubin model} \/\cite{Rubin}. \ \ \ The
particle-impurity with the mass $M$ is located in an infinite,
$1D$ harmonic lattice, particles of which have the same mass $m$
($K$ is the coupling constant). The crucial quantity here is the
mass ratio, $q=M/m \geq 1$. In case of $q \gg 1$ the movement of a
particle with mass $M$ is the same as the stochastic movement of a
Brownian particle (the memory-free limit) \cite{Zwanzig_book}. If
the mass ratio $q=1$, then the system consists of particles of the
same mass, that corresponds to the model for acoustic phonons in
crystalline solids at ordinary temperatures (the strong memory
limit) \cite{Lee4}. Thus, the Rubin model contains both
memory-free and strong memory limits, transition to which is
defined by value $q$. The term $\widetilde{a}(z)$ and the
frequency parameter $\omega^{(2)}$ have the following forms here
\cite{Zwanzig_book}: \be \widetilde{a}(z)=
\frac{q}{(q-1)z+\sqrt{z^2+4K/m}},\ \ \ \omega^{(2)}=2K/M.
\label{LVACF} \ee From last expression we find \be
\widetilde{a}(z=0)=q\sqrt{\frac{m}{4K}}, \ \ \
\widetilde{a}'(z=0)=-\frac{M}{4K}(q-1). \ee Then, from Eq.
\Ref{nM} we find the expression \be \delta=\frac{q}{2}|q-1|,
\label{rubin} \ee related 
$\delta$ with the quantity
$q$. It is seen from the given relation that for the strong memory
limit at $q=1$ we have $\delta=0$, and for the memory-free one we
obtain $\delta \to \infty$. The squared $q$-dependence in Eq.
\Ref{rubin} defines the transition in this model from one limit to
the other.

B. \textit{Ideal gas.} \ \ \ TCF of density fluctuations for an
ideal gas \cite{Resibua} is \be \phi(t)=\frac{\langle
\rho(0)^*\rho(t) \rangle}{\langle |\rho(0)|^2
\rangle}=\exp(-\Omega_1^2 t^2/2). \label{Gauss} \ee Note that the
Laplace transform of $\phi(t)$ is related to the Laplace transform
of autocorrelator for a longitudinal component of velocity of
local density fluctuations, $a_l(t)$, by \be
\widetilde{\phi}(z)=[z+\Omega_1^2 \widetilde{a}_l(z)]^{-1}.
\label{cont} \ee The term $a_l(t)$ is similar to VACF and its
frequency moment $\omega^{(2)}$ is related to $\Omega_1^2$ by
$\omega^{(2)}=2\Omega_1^2$ \cite{Balucani_Zoppi}. Moreover,
applying the Laplace transform to Eq. \Ref{Gauss} and taking into
account Eq. \Ref{cont} one finds \bn \widetilde{a}_l(z)=
\sqrt{\frac{2}{\Omega_1^2 \pi}}
\exp[-z^2/(2\Omega_1^2)]\;\textrm{erfc}^{-1}\left
(\frac{z}{\sqrt{2\Omega_1^2}}\right ) -\frac{z}{\Omega_1^2}
\nonumber \en with the following low-frequency properties \be
\lim_{z \to 0} \widetilde{a}_l(z)^2 =\frac{2}{\pi\omega^{(2)}},\ \
\ \lim_{z \to 0}
\widetilde{a}_l'(z)=\frac{2-\pi}{\pi\omega^{(2)}}. \ee Finally,
one finds from Eq. \Ref{nM} that \be \delta
=\frac{4(\pi-2)}{\pi(4-\pi)} \approx 1.69. \label{id_gas} \ee It
is the evident that the relaxation process related to fluctuations
of a longitudinal velocity (momentum) component in an ideal gas is
characterized by pronounced memory and independent from model
parameters such as $\omega^{(2)}$.

C. \textit{Interacting $2D$ electron gas at long wavelengths and
at $T=0$.} In Ref. \cite{Lee4} it was shown that TCF of density
fluctuations and TCF of a longitudinal momentum component for this
model are \bn \phi(t)=J_{0}(2\sqrt{\Omega_1^2}t), \ \ \
{a}_l(t)=\frac{J_1(2\sqrt{\Omega_1^2}t)}{\sqrt{\Omega_1^2}t},
\label{J_1} \en where $\Omega_1^2=k^2 \epsilon_F^2$,
$\epsilon_F^2$ is the Fermi energy and $k$ is measured in units of
the Fermi vector $k_F$, $J_n$ is the Bessel function of the $n$th
order. It is necessary to note that in the case of such time
dependence of $a_l(t)$, this TCF and its memory function have the
same time behavior. As a result, they have the same relaxation
time scales and $\delta=1$ that is easily obtained by the Laplace
transform of Eq. \Ref{J_1} with taking into account Eq. \Ref{nM}.

\textit{Long time tails.} \ \ \ As known, TCF's in various
physical systems can be characterized by long lasting tails
\cite{Bei}. For example, the long time behavior of VACF in liquids
\cite{Zwanzig_book} is \be \lim_{t \to \infty} a(t) \sim
t^{-\frac{d}{2}}, \label{llt}\ee where $d=2$ and $3$ are
two-dimensional and three-dimensional cases, correspondingly. In
accordance with the Tauberian theorem, this is equivalent to \be
\lim_{z \to 0} \widetilde{a}(z) \sim z^{\frac{d}{2}-1}. \ee Then,
one obtains the following low-frequency properties
\be
\left\{
\begin{array}{lcl}
\displaystyle{\widetilde{a}(z)^2 \sim 1,\ \widetilde{a}'(z)=0},  &\quad{\rm if}& d=2,\\
\displaystyle{\widetilde{a}(z)^2 \sim z,\ \widetilde{a}'(z)\sim\frac{1}{\sqrt{z}}},  &\quad{\rm if}& d=3,\\
\end{array}
\right.  \label{sys}
\ee
and from Eq. \Ref{nM} we find that
$\delta=0$ for both cases. It is evident, that the long time
dynamics related to the fractal law \Ref{llt} is characterized by
memory effects which appear directly by conserved long-living
correlations. From this point of view, the source of such behavior
is in the presence of arbitrary fragments of the regular motion in
the system dynamics likely to be found in the billiard model
\cite{Viva}.

\textit{Anomalous diffusion.} Let us consider anomalous diffusion
phenomena on an example of a free particle coupled to a fractal
heat bath -- the model used to study both sub- and superdiffusion.
We don't consider the superdiffusion region including ballistic
limit with $\alpha \geq 2$ \cite{sev}. The VACF for this model can
be written as \be a(t)=E_{2-\alpha}(-\gamma_{\alpha}
t^{2-\alpha}), \ \gamma_{\alpha}=\frac{\pi
A_0}{mk_BT\sin(\alpha\pi/2)}, \label{anom_d} \ee where
$E_{\alpha}(t)$ is the Mittag-Leffler function, $A_0$ is the
strength of the coupling \cite{Lutz}. Ordinary diffusion
corresponds to $\alpha=1$, superdiffusion is associated with
motion at $1<\alpha<2$, while subdiffusion occurs when
$0<\alpha<1$. Eq. \Ref{anom_d} for $\alpha\neq 1$ has an inverse
power-law tail at long times \cite{Lutz2} \be
a(t)\sim\frac{t^{\alpha-2}}{\gamma_{\alpha}}. \label{long_tail}
\ee One can readily deduce  that for subdiffusion with $\alpha<1$
the parameter $\delta \to 0$. For example, at $\alpha=1/2$ we
obtain the same result as for a long time tail with $d=3$ [see Eq.
\Ref{sys}]. Let us next consider superdiffusion with $1<\alpha<2$.
Then, for $\alpha=3/2$ we have \be \lim_{z \to 0}
\widetilde{a}(z)^2 \sim z^{-1}, \ \ \ \lim_{z \to 0}
\widetilde{a}'(z) \sim z^{-\frac{3}{2}}. \label{superd} \ee Taking
into account Eq. \Ref{superd} we find from Eq. \Ref{nM} that
$\delta \to \infty$. As a result, although the VACF behavior in
this case strongly deviates from the exponential relaxation
pattern, its memory function decays rather fast in comparison with
the VACF, i.e. superdiffusion dynamics is similar to an ordinary
Markovian process \cite{vosem}.

In conclusion, the notion of strong dynamical memory  is rather
characteristic for the  statistical properties of the time evolution
of many systems, and as demonstrated with our study it depends on
physical features of the underlying process. From the point of view
of a recurrent relation method, properties of memory can be
associated with the topology of a Hilbert space, where a dynamical
variable is represented as a vector \cite{Lee1}. Memory effects then
occur  both within a finite-dimensional space, and in an
infinite-dimensional one, where the VACF is dissipative. On the
basis of the GLE formalism we propose a convenient measure of memory
effects in dynamical systems. This quantity is based on a definition
that allows one to extract in a tractable manner the role of memory
behavior for the dynamics, and moreover, allows one to test the
validity of memory-free approximations. The simplicity of the given
measure makes its application possible in the analysis of various
complex systems. Although the role of memory effects for chaotic
processes is not yet settled, the measure $\delta$ may provide a
useful guide in establishing an interrelation of manifest memory for
such fundamental properties as chaotic behavior \cite{FVMP} and
ergodicity \cite{last}.

A.V.M. acknowledges L. Rondoni, B. Spagnolo and N. Lemke for the
useful correspondence. This work was partially supported by the
Russian Ministry of Education and Science (Grant No. 03-06000218a)
and RFBR (Grants No. 05-02-16632, No. 03-02-96250).



\begin{thebibliography}{99}

\bibitem{Zwanzig_book} R. Zwanzig, \textit{Nonequilibrium statistical
mechanics} (University Press, Oxford, 2001);
H. Mori, Rep. Prog. Phys. \textbf{34}, 399 (1965).

\bibitem{Lee1} M.H. Lee, Phys. Rev. E \textbf{62}, 1769 (2000),
M.H. Lee, Phys. Rev. Lett. \textbf{85}, 2422 (2000).

\bibitem{Hanggi} P. H{\"a}nggi, P. Talkner, and M. Borkovec, Rev. Mod. Phys. \textbf{62}, 251
(1990); S.C. Kou, X.S. Xie, Phys. Rev. Lett. \textbf{93}, 180603
(2004).

\bibitem{GHT} H. Grabert, P. H\"anggi, and P. Talkner, J. Stat. Phys. {\textbf 22},
537 (1980).

\bibitem{Lee0} M.H. Lee, Phys. Rev. Lett. \textbf{51}, 1227
(1983).

\bibitem{Balucani_Zoppi}  U. Balucani and M. Zoppi, {\it Dynamics of the Liquid State}
(Clarendon Press, Oxford, 1994); X. Xia, and P. G. Wolynes, Phys.
Rev. Lett. \textbf{86}, 5526 (2001); A.V. Mokshin \textit{et al}.,
J. Chem. Phys. \textbf{121}, 7341 (2004); R.M. Yulmetyev
\textit{et al}.,
Phys. Rev. E \textbf{68}, 051201 (2003).

\bibitem{Ben}
P. Licinio, and M. B. L. Santos, Phys. Rev. E. \textbf{65}, 031714
(2002).

\bibitem{Fer} J.L. Ferreira, G.O. Ludwig, and A. Montes, Plasma Phys. Controlled Fusion
\textbf{33}, 297 (1991).

\bibitem{Vain} M.H. Vainstein, D.A. Stariolo, and J.J. Arenzon, J. Phys. A: Math. Gen. \textbf{36}, 10907 (2003).

\bibitem{Peyr} M. Peyrard, Phys. Rev. E \textbf{64}, 011109
(2001).

\bibitem{Bouch} J.P. Bouchaud \textit{et al}.,
Phys. Rev. Lett. \textbf{67}, 3840 (1991).

\bibitem{HT} P. H\"anggi, and P. Talkner, Phys. Rev. Lett.
\textbf{51}, 2242 (1983).

\bibitem{Boon}  J.P. Boon and S. Yip,
\textit{Molecular Hydrodynamics} (McGraw-Hill, New-York, 1980);
J.P. Hansen and I.R. McDonald,
\textit{Theory of Simple Liquids} (Academic Press, London, 1986).

\bibitem{sm} Nevertheless, it should be kept in mind, that a large value for $\delta$ may as well mimic
an approximate Markovian \textit{nonlinear} process (driven by
white noises) with  the  linear (Mori-like) GLE possessing strong
memory effectively describing a nonlinear (Kawasaki-like) GLE
dynamics with small memory effects.

\bibitem{Resibua} P. Resibois and M. De Leener,
\textit{Classical Kinetic Theory of Fluids} (Wiley, New-York,
1977).

\bibitem{Langevin} M.P. Langevin, Comptes Rendus \textbf{146}, 530
(1908).

\bibitem{Rubin} R.~J. Rubin, J. Math. Phys. {\bf 1}, 309 (1960); H. Nakazawa, Prog.
Theor. Phys. Suppl.  {\bf 36}, 172 (1966).

\bibitem{Lee4} M.H. Lee, Journal of Molecular Structure (Theochem)
\textbf{336}, 269 (1995).

\bibitem{Bei} H. Van Beijeren, Rev. Mod. Phys. \textbf{54}, 195
(1982).

\bibitem{Viva} F. Vivaldi \textit{et al}.,
Phys. Rev. Lett. \textbf{51}, 727 (1983).

\bibitem{sev} In work by R. Morgado et al.,
Phys. Rev. Lett.
\textbf{89}, 100601 (2002) it was shown that the restriction to
this $\alpha-$region guarantes the fluctuation-dissipation theorem
fulfilment.

\bibitem{Lutz} E. Lutz, Phys. Rev. E \textbf{64}, 051106 (2001).

\bibitem{Lutz2} E. Lutz, Europhys. Lett. \textbf{54}, 293 (2001).

\bibitem{vosem} From Eq. \Ref{GLE_3} and
Eq. \Ref{long_tail} at $\alpha=3/2$ one obtains that $\lim_{t \to
\infty} a(t) \sim t^{-1/2}$ and $\lim_{t \to \infty} M_1(t) \sim
t^{-2}$.

\bibitem{FVMP} M. Falcioni, A. Vulpiani, G. Mantica, and S.
Pigolotti, Phys. Rev. Lett. \textbf{91}, 044101 (2003).

\bibitem{last} Corresponding characteristics of chaoticity and
ergodicity can also be expressed in terms of TCF's. For example,
it was shown in Ref. \cite{Lee1} that $\widetilde{a}(z=0)$
characterizes the ergodic properties of a system: if
$0<\widetilde{a}(z=0)<\infty$, a system is ergodic, and if
$\widetilde{a}(z=0)=0$ or $\infty$, it is not. As for chaoticity,
some interrelations between entropy and diffusion coefficients
[and, consequently, the VACF, i.e.
$\widetilde{a}(z=0)=Dm/(k_{B}T)$] were recently suggested in [M.
Dzugutov, Nature (London) \textbf{381}, 137 (1996);
S. Bastea, Phys. Rev. E \textbf{68}, 031204 (2003); J.-L.
Bretonnet, J. Chem. Phys. \textbf{117}, 9370 (2002)].

\end{thebibliography}
\end{document}